\documentclass[aps,prl,twocolumn,amsfonts,showpacs]{revtex4}
\usepackage{epsfig,amsopn}
\def\bxi{\mbox{\boldmath{$\xi$}}}
\begin{document}
\date{\today}
%%%%%%%%%%%%%%%%%%%%%%%%%%%%%%%%%%%%%%%%%
\title{Diffusion of Asymmetric Swimmers}
\author{Andrew D. Rutenberg}
\author{Andrew J. Richardson}
\author{Claire J. Montgomery}
\affiliation{Department of Physics and Atmospheric Science, 
Dalhousie University, Halifax, Nova Scotia, Canada B3H 3J5}
% 05 statistical physics
% 05.40.-a fluctuation phenomena
% 05.40.Fb Random walks and Levy flights
% 87 Biological and Medical Physics
% 87.16-b Subcellular structure and processes
% 87.16.Dg Membranes, bilayers, and vesicles
%%%%%%%%%%%%%%%%%%%%%%%%%%%%%%%%%%%%%%%%%%%%%%%%%
\begin{abstract}
Particles moving along curved trajectories will diffuse if the curvature
fluctuates sufficiently in either magnitude or orientation.  We consider
particles moving at a constant speed with either a fixed 
or with a Gaussian distributed curvature magnitude. At small 
speeds the diffusivity is independent of the speed.  At larger 
particle speeds, the diffusivity depends on the speed through a novel exponent.  
We apply our results to intracellular transport of vesicles.  In sharp contrast 
to thermal diffusion, the effective diffusivity {\em increases} with vesicle 
size and so may provide an effective means of intracellular transport. 
\end{abstract}
\maketitle
%%%%%%%%%%%%%%%%%%%%%%%%%%%%%%%%%%%%%%%%%%%%%%%%%%%%%%%%%

The thermal Stokes-Einstein diffusivity of a sphere
decreases as the particle radius $R$ increases \cite{Berg93}.  For
this reason, while diffusive transport is used for 
individual molecules within living cells \cite{Alberts2002}, larger objects
such as vesicles and pathogens often use active means of transport. 
While many intracellular vesicles appear to be
transported by molecular motors directed along existing cytoskeletal tracks 
\cite{motorvesicle,Alberts2002}, {\em undirected} actin-polymerization mediated 
vesicle transport has been reported in some endosomes, lysosomes, other endogenous 
vesicles, and phagosomes \cite{vesicles,reviews}.  
Active transport is also observed in the actin-polymerization-ratchet motility 
of certain bacteria \cite{reviews,rickettsiae} and virus particles \cite{vaccinia} 
within host cells.  It is important to characterize the transport properties
of vesicles that are not moving along pre-existing cytoskeletal tracks. 

Existing discussions of the motion of actively propelled microscopic
particles, or ``swimmers'', assumes that in the absence of thermal fluctuations
particles would move in straight trajectories \cite{Lovely75,Berg93}.  
Thermal rotational diffusion will then randomly re-orient the trajectory \cite{Berg93}, 
so that over long times diffusive transport will be observed.  However in 
actin-polymerization based motility, particles appear to be attached to their 
long actin tails \cite{Gerbal2000} which in turn are embedded in the cytoskeleton \cite{Theriot92}.  
While thermal fluctuations will thereby be severely reduced, the
actin-polymerization itself is a stochastic process with its own fluctuations \cite{Alberts2002,Mogilner96}.
These intrinsic fluctuations can explain the observed curved trajectories, as well
as the variation of the curvature over time \cite{Rutenberg2001}.  The diffusivity of
such asymmetrically moving particles has not been previously explored. 

In this letter, we study asymmetric swimmers that would
move in perfect circles in the absence of fluctuations.  We examine
both a ``broken swimmer'' with a fixed curvature magnitude and an axis of curvature that is re-oriented by
fluctuations (rotating curvature, RC), and a ``microscopic swimmer'' with a 
normally-distributed curvature that is spontaneously generated by fluctuations (Gaussian curvature, 
GC).  In both of these systems, fluctuations lead to diffusion at long times.   
We use computer simulations to measure  the diffusivity of these systems as a function of 
the root-mean-squared curvature $K_0$, the particle speed $v$, and the timescale characterizing
the curvature dynamics $\tau$.  

We obtain some exact results from polymer systems, where each polymer configuration 
represents a possible particle trajectory. Indeed, a broken swimmer with
a fixed curvature magnitude in $d=3$ is exactly analogous to the
hindered jointed chain discussed by Flory \cite{Flory89} and we thereby recover the entire
scaling function exactly.  In that case, the diffusivity is independent of 
particle speed $v$.  For Gaussian curvatures and for systems in restricted geometries ($d=2$), 
the polymer analogy gives us the diffusivity only in the limit of slow speeds. 
At larger speeds, our simulations over $5$ decades of speed show that 
diffusivity depends on particle speed with a non-trivial exponent $\lambda$. 
The diffusivity appears to be dominated by the occasional long straight segments of trajectory that 
occur when the curvature is small. Scaling arguments based on this observation are
consistent with the measured exponent $\lambda_{2d} =0.98 \pm 0.02$ in $d=2$, but do not recover 
our measured exponent $\lambda_{3d}=0.71 \pm 0.01$ in $d=3$. 

A curved path has a curvature magnitude $K
\equiv 1/R$, where $R$ is the instantaneous radius of curvature. If we
describe a particle trajectory by a position ${\bf r}(t)$, then the vector
curvature is defined by the cross-product
$ {\bf K} \equiv { {\bf v} \times \dot{\bf v} / v^3}$,
where $v= |{\bf v}|$ is the speed and the dot $\dot{}$ indicates a time-derivative.  
For uniform motion around a circle, $R$ is the radius of the circle, and ${\bf K}$ is
oriented perpendicular to the circle along the axis.
We consider particles moving at a constant speed and with an 
instantaneous curvature ${\bf K}$, so that
$\dot{\bf r} = {\bf v}$ and $ \dot{\bf v} = - v {\bf v} \times {\bf K}$.
For ``rotating curvature'' dynamics (RC) we
fix the curvature magnitude $|{\bf K}|=K_0$ but allow the
curvature to randomly rotate around the direction of motion:
\begin{equation}
	\dot{\bf K}_{RC} = \xi \hat{\bf v} \times {\bf K}
\label{EQN:RCkdyn}
\end{equation}
where the unit-vector $\hat{\bf v}= {\bf v}/v$, 
the Gaussian noise $\xi$ has zero mean, and $\left< \xi(t) \xi(t') \right> = 2 
\delta(t-t')/ \tau$ with a characteristic timescale $\tau$.  
This represents the simplest description of a mesoscopic swimmer that has a ``locked-in''
curvature due to, e.g., an asymmetric shape.  For ``Gaussian curvature'' dynamics (GC) the curvature
magnitude changes as well: 
\begin{equation}
	\dot{\bf K}_{GC} = -{\bf K}/\tau+\bxi
\label{EQN:KGC}
\end{equation}
where the noise $\bxi$ is perpendicular to ${\bf v}$ with zero mean
and $\left<\bxi(t) \cdot \bxi(t')\right>= \delta(t-t') K_0^2/\tau$, such
that $\left<{\bf K}^2\right> = K_0^2$.  
This represents the simplest description of a microscopic swimmer ``trying'' to swim in 
a straight line subject to intrinsic fluctuations in the motion.  The resulting curvatures are 
Gaussian distributed in each component. 
For particles restricted to two-dimensions with either RC or GC dynamics,
we only use the normal ($\hat{z}$) component of the vector-curvature to
update the velocity within the plane, i.e. 
$ \dot{\bf v} = - v {\bf v} \times \hat{\bf z} K_z$
in $d=2$. 

There are two natural timescales. We explicitly introduce $\tau$, which
controls the noise correlation and so sets the timescale over which the curvature
changes.  There is also the inverse of the angular rotation rate, $t_c \equiv 1/(v K_0)$. 
Diffusion will only be observed for elapsed times $t$ much greater than any other 
timescale in the system, i.e. $t \gg t_c$ {\em and} $t \gg \tau$.  
The diffusivity of a particle is given by 
$D \equiv \left<r^2\right>/(2dt)$ in the limit as the elapsed time 
$t \rightarrow \infty$, in spatial dimension $d$.  

A polymer chain with fixed bond lengths ($\ell$) and angles ($\theta_f$), 
and with independent bond rotation potentials ($V(\phi_f)$) \cite{Flory89} is statistically
identical to the continuous RC trajectory in $3d$
if for a discrete time-step $\Delta t$ we take $\ell =  v \Delta t$.
The end-to-end distance for a long $n$-bond polymer is 
$\left<r^2\right> = n  \ell^2 C_n$.  The correspondence is complete
as the elapsed time $t = n \Delta t \rightarrow \infty$.   
The bond and dihedral angles determine
$C_\infty = (1 + \cos\theta_f) (1+\left<\cos \phi_f\right>) / [(1- \cos \theta_f)
(1- \left<\cos \phi_f \right>)]$ \cite{Flory89}.  
Swimmers follow continuous
paths, so we take the limit of small $\Delta t$ and 
fix the polymer rotation angle from the curvature  in that limit
by $\theta_f = K_0 v \Delta t$, and rotate the curvature by 
$< \phi_f^2 >  = 2 \Delta t/\tau$ in agreement with Eqn.~\ref{EQN:RCkdyn}.
In the limit $\Delta t \rightarrow 0$ we recover the {\em exact} result 
$D = 1/(3 K_0^2 \tau)$ in $d=3$. Remarkably, $D$ is independent of $v$. 

For a GC trajectory in $d=3$, there is no obvious polymer analogy 
since the curvature magnitude evolves with time.  
In the limit of $\tau \rightarrow 0$ however, the curvature is independently
Gaussian distributed at every point along the trajectory and the diffusivity 
can be extracted from the ``worm-like chain'' polymer model originally 
solved by Kratky and Porod \cite{Doi}.  The result is $D = 1/(3 K_0^2 \tau)$
in the limit of small $\tau$. Note that the diffusivity diverges as 
$1/\tau$ so this is the leading asymptotic dependence for small $\tau$. 
We obtain the same diffusivity for both RC and GC dynamics for small $\tau$. 

We use these exact results to define natural dimensionless scaling functions for 
the diffusivity of microscopic swimmers:
\begin{equation}
	\tilde{D}_{Rd, Gd}(\tilde{v}) \equiv  D K_0^2 \tau
\label{EQN:RCGC}
\end{equation}
where the index $Rd$ {\em or} $Gd$ indicates both the dynamics (RC or GC) 
and the spatial dimensionality $d$, and
\begin{equation}
	\tilde{v} \equiv v K_0 \tau
\end{equation} 
is a dimensionless speed.  In terms of these scaling functions we have 
$\tilde{D}_{R3}(\tilde{v})=\tilde{D}_{G3}(0)= 1/3$.
The same Kratky-Porod approach in $d=2$ gives $\tilde{D}_{R2}(0)=\tilde{D}_{G2}(0)= 1$.

\begin{figure}[h]
	\centerline{\epsfxsize = 1.6truein \epsfbox{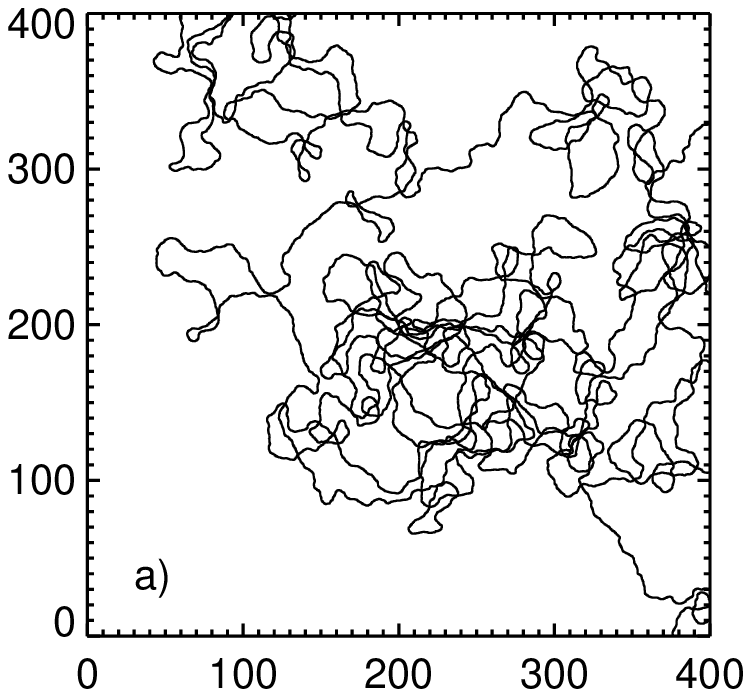} 
	\epsfxsize=1.6truein \epsfbox{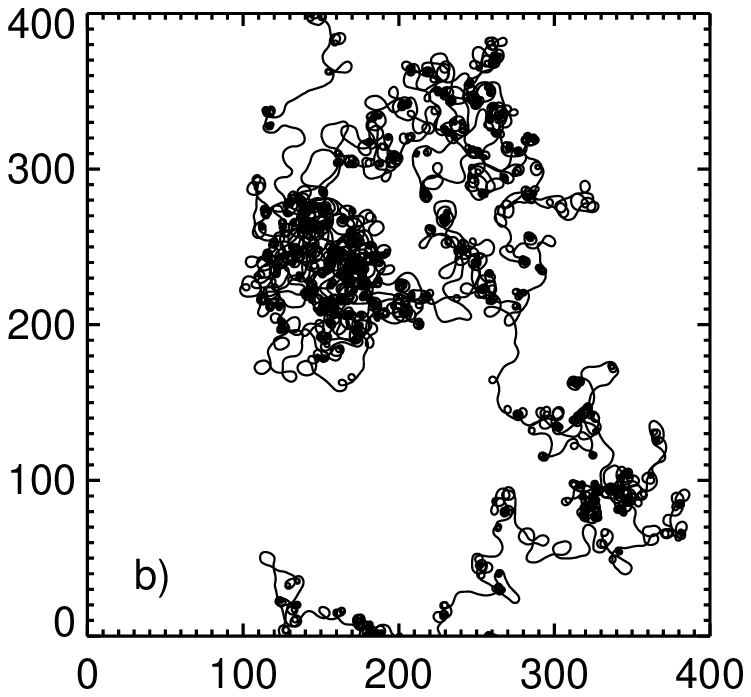}}
\vspace*{-0.2in}
	\caption{a) Particle trajectory with GC dynamics in $d=2$ with $\tilde{v}=0.1$. 
The particle does not complete a circular loop before $\bf K$ changes significantly.
b) With $\tilde{v}=100$. The particle can complete
many circular loop before $\bf K$ changes, however straight segments are seen when 
$|{\bf K}|$ is small.  The result is a characteristic ``knotty wool'' appearance. 
In both cases $K_0=1$.}
	\label{FIG:traj}
\end{figure}

We have simulated the trajectories of large-numbers of independent particles with 
RC and with GC dynamics.  For fixed $v$ and $K_0$, 
we varied $\tau$ to explore the scaled velocity $\tilde{v} \equiv v K_0 \tau$ over $5$ orders
of magnitude.  For each $\tilde{v}$, we averaged over the trajectories of at least $1000$ particles.
We explicitly integrated the dynamical equations using a simple Euler update with a small timestep $\Delta t$. 
In all cases $t \gg  \tau \gg \Delta t$ and $t \gg t_c \gg \Delta t$, with separation of 
timescales by factors of $10-100$. Systematic errors due to 
$\Delta t$ and $t$ are below our noise levels, and statistical errors (when not shown) 
are smaller than the size of our plotted points.  Consistently, our numerical results agree
with all exact results from the polymer analogy.  We illustrate the trajectories
that we observe in $d=2$ in Fig.~\ref{FIG:traj}, with both small and large scaled speeds $\tilde{v}$.
In both cases the curvature $K_0=1$, but particles only complete loops at
large $\tilde{v}$. Qualitatively similar trajectories are seen in $d=3$ with GC curvature dynamics. 

\begin{figure}
	\centerline{\epsfxsize = 3.0truein
	\epsfbox{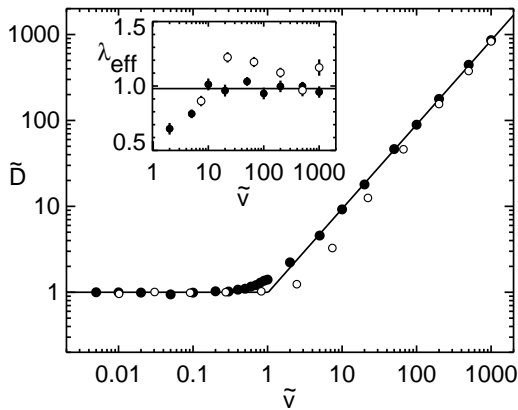}}
\caption{Dimensionless diffusivities $\tilde{D} \equiv D K_0^2 \tau$ for rotating 
(open circles, $\tilde{D}_{R2}$) 
and Gaussian (filled circles, $\tilde{D}_{G2}$) curvature dynamics in $d=2$, 
plotted against dimensionless particle speed $\tilde{v} \equiv v K_0 \tau$.   
Also shown with solid lines are the small $\tilde{v}$ 
asymptote $\tilde{D}=1$ and the large $\tilde{v}$ best-fit asymptote 
$\tilde{D} \sim \tilde{v}^{\lambda_{2d}}$. 
The inset shows the effective exponents, with the solid line indicating the best-fit
$\lambda_{2d}=0.98 \pm 0.02$. 
\label{FIG:2d}}
\end{figure}

In $d=2$, shown in Fig.~\ref{FIG:2d}, 
both rotating curvature (open circles) and Gaussian curvature (filled circles) approach 
their asymptotic value of $\tilde{D}_{R2}(0)=\tilde{D}_{G2}(0) =1$ at small $\tilde{v}$. 
At $\tilde{v} \approx 1$ there is a sharp cross-over to a large-$\tilde{v}$ power-law regime, 
characterized by an exponent $\lambda_{2d}$ where 
$\tilde{D}_{R2} \sim \tilde{D}_{G2} \sim \tilde{v}^{\lambda_{2d}}$
for large $\tilde{v}$.  We show the effective exponents 
$\lambda_{eff} \equiv \Delta \log{(D K_0^2 \tau)}/\Delta \log{\tilde{v}}$ 
between consecutive points in the inset of Fig.~\ref{FIG:2d}, as well as the best fit exponent
$\lambda_{2d} =0.98 \pm 0.02$.  We fit $\lambda_{2d}$ from the large-$\tilde{v}$ GC data only, 
due to the systematic cross-over remaining in the RC data even at large $\tilde{v}$.  

\begin{figure}
	\centerline{\epsfxsize = 3.0truein
	\epsfbox{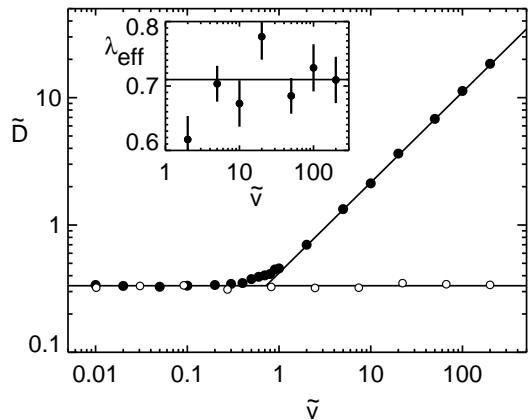}}
\caption{Dimensionless diffusivities $\tilde{D} \equiv D K_0^2 \tau$ 
for rotating (open circles, $\tilde{D}_{R3}$) 
and Gaussian (filled circles, $\tilde{D}_{G3}$) curvature dynamics in $d=3$, plotted against 
dimensionless particle speed $\tilde{v} \equiv v K_0 \tau$.   Solid lines show the 
exact result $\tilde{D}_{R3}=1/3$, as well as the large $\tilde{v}$ power-law 
asymptote $\tilde{D}_{G3} \sim \tilde{v}^{\lambda_{3d}}$. 
The inset shows effective exponents between sequential
points, with a solid line indicating the best-fit $\lambda_{3d}=0.71 \pm 0.01$.  
\label{FIG:3d}}
\end{figure}

Simulations in $d=3$ with rotating curvature (RC) dynamics leads to a diffusivity in excellent
agreement with the exact result from polymer physics, $\tilde{D}_{R3} = 1/3$, as shown by open circles
in Fig.~\ref{FIG:3d}.  Gaussian curvature dynamics ($\tilde{D}_{G3}$, filled circles) 
has the same behavior for small $\tilde{v}$, but exhibits a sharp crossover at $\tilde{v} \approx 1$ 
to a power-law
regime $\tilde{D}_{G3} \sim \tilde{v}^{\lambda_{3d}}$ for large $\tilde{v}$. We find the best-fit exponent
is $\lambda_{3d}=0.71 \pm 0.01$, as shown by the solid line in the inset of Fig.~\ref{FIG:3d}. 
Because $\lambda_{3d}<1$, this scaling curve may be used to uniquely identify the dynamical timescale $\tau$
if $D$, $K_0$, and $v$ are measured experimentally. 

Is there a simple way of understanding the asymptotic behavior of $\tilde{D}$? 
For RC dynamics in $d=3$ the instantaneous curvature does not change in magnitude even while the 
curvature axis wanders.  The particle will go in a circular trajectory, not contributing to 
diffusivity, until the curvature axis wanders significantly. The result is
a random walk with step size given by the radius of curvature
$\Delta r \sim 1/K_0$ and an interval between steps of $\tau$, 
leading to $D \sim 1/(K_0^2 \tau)$. This qualitatively explains why
the exact result $\tilde{D}_{R3}=1/3$ is independent of $\tilde{v}$.  

It is more difficult to understand the 
$\tilde{D} \sim \tilde{v}^\lambda$ behavior for large $\tilde{v}$ in 
the other systems.  We start with a simple scaling argument 
based on the assumption that the relatively straight segments shown in Fig.~\ref{FIG:traj} b) 
dominate the diffusivity.  The interval between periods of small curvature should
be on order the autocorrelation time $\tau$.  The length $\Delta r$ of the straight segments
are determined by how long the interval of small curvature lasts, $\Delta t$, 
since $\Delta r \approx v \Delta t$.  For the segment to be straight, the
curvature must be less than the inverse length, i.e.  $K_{max} \lesssim 1/\Delta r$. 
The fraction of the time we have small curvature below $K_{max}$ in magnitude
should be proportional to the probability of having curvature below $K_{max}$.
In $d=2$ only the normal component of curvature affects the dynamics, so that
$P(K) \approx {\rm const}$ for $K \ll K_0$. This applies both to GC and RC. We therefore
expect $\Delta t \sim \tau K_{max}/K_0$. We maximize $K_{max}$ 
to maximize the contribution to $D \approx \Delta r^2/\tau$ and find 
$\tilde{D}_{G2} \sim \tilde{D}_{R2} \sim \tilde{v} $ as $\tilde{v} \rightarrow \infty$.
This indicates that $\lambda_{2d}=1$, which is 
consistent with our best-fit value $\lambda_{2d}=0.98 \pm 0.02$.  
However, in $d=3$ for GC dynamics the same argument leads to $\lambda_{3d}=2/3$ since two Gaussian
distributed components of the curvature gives
$P_<(K_{max}) = \int_0^{K_{max}} dK P(K) \sim K_{max}^2/K_0^2$ for $K_{max} \ll K_0$. 
This is inconsistent with our measured value of $\lambda_{3d}=0.71 \pm 0.01$, 
with a significant $4 \sigma$ variation.  

At what radius $R_c$ does a small spherical particle achieve a higher diffusivity by actively 
swimming, as compared to passive thermal diffusion characterized by $D_T = k_B T/ (6 \pi \eta R)$ \cite{Berg93}? 
We can answer this question within the context of actin-polymerization based motility
of small intracellular particles, since the size dependence of $K_0$, $v$, and $\tau$ is 
known, at least approximately. 
With the approximation that $n$ propulsive actin filaments are randomly distributed
over a particle of size $R$, the curvature of the trajectory will be $K_0 \propto 1/(R\sqrt{n})$
\cite{Rutenberg2001}.  With a size-independent surface-density of 
filaments we obtain $K_0 \approx A/R^2$, with a constant of 
proportionality $A$.  By observations of {\em Listeria monocytogenes} 
we estimate $A \approx 0.1 \mu m$ 
\cite{Rutenberg2001}.  We also {\em conservatively} assume size-independent values for cytoplasmic 
viscosity $\eta \approx 3 Pa \cdot s$ \cite{Berg93,Rutenberg2001}, speed $v \approx 0.1 \mu m/s$, 
and autocorrelation decay time $\tau \approx 100 s$ \cite{Sechi97}.
We find that the micron-scale bacterium {\em L. monocytogenes} has $\tilde{v} \approx 1$, 
so that smaller particles will have $\tilde{v} >1$. 
Using the large $\tilde{v}$ asymptotic behavior of $\tilde{D}_{G3}$ shown in Fig.~\ref{FIG:3d}, 
$D \approx 0.41 \tilde{v}^{\lambda_{3d}}/(K_0^2 \tau)$, and the 
size-dependence $K_0 \approx A/R^2$, we obtain 
\begin{equation}
	D_{G3} \sim R^{4-2 \lambda_{3d}} v^{\lambda_{3d}}/(A^{2-\lambda_{3d}} \tau^{1-\lambda_{3d}}),
	\label{EQN:G3size}
\end{equation}
with a measured $\lambda_{3d}=0.71 \pm 0.01$.  In dramatic contrast to thermal diffusion,  
$D$ {\em increases} with increasing particle size.  Comparing with $D_T$ we find that for all sizes 
{\em above} $R_c \approx 80 nm$ a particle will have a higher diffusivity
by actively swimming by the actin-polymerization mechanism than by passive thermal diffusion.
Provocatively, this is in the middle of the vesicle-size distribution seen in neural 
systems \cite{vesiclesize}.

Our treatment of microscopic swimmers has ignored thermal fluctuations.
A ``rocket'' traveling straight at speed $v$ 
that is re-oriented only by thermal effects will have 
$D_u=4 \pi \eta R^3 v^2/(3 k_B T)$ \cite{Berg93}.  In comparison with our results for $D$,
we find that $D<D_u$ for particles larger than $R_u \approx 0.07 nm$.  For actin-polymerization 
based motility, intrinsic fluctuations appear to dominate thermal fluctuations at the 
particle sizes where active transport is advantageous.   

In summary, we  find that diffusivities of asymmetric microscopic swimmers depends on 
whether the swimmers are restricted to $2d$ or $3d$, and whether they have fixed asymmetries (RC)
or the asymmetries are spontaneously generated (GC). 
Diffusivities are independent of particle speed at low speeds, in agreement with analogous
polymer systems.  At higher speeds an anomalously large diffusivity is observed that depends on the
particle speed by $\tilde{v}^\lambda$ where $\lambda_{2d}=0.98 \pm 0.02$, 
in agreement with a scaling argument for $\lambda_{2d}=1$. 
However $\lambda_{3d}=0.71 \pm 0.01$, which significantly differs from our simple scaling result in $d=3$.  
We apply our results to intracellular bacteria, virus particles, and vesicles that move via
actin-polymerization. We find that diffusivities due to asymmetric swimming exceed thermal
diffusivities for particles {\em larger} than approximately $80 nm$. As a result asymmetric swimming
may provide a viable intracellular transport mechanism even for vesicle-sized particles.  
We find that for the relevant dynamics (GC in $d=3$), diffusivities should increase with 
particle size, speed, and filament turnover rate, and also with smaller curvatures for a given size.   It 
is interesting that the bacterium {\em Rickettsiae rickettsii} exhibits actin-polymerization
intracellular motility with smaller intra-cellular speeds but straighter trajectories
\cite{rickettsiae,Rutenberg2001} --- raising the question of whether maximal diffusivity is selected for 
in this or other biological systems.

This work was supported financially by an NSERC discovery grant. 
C. Montgomery would also like to acknowledge support from an NSERC USRA.

%%%%%%%%%%%%%%%%%%%%% REFERENCES %%%%%%%%%%%%%%%%%%%%%%%%%%%%%%%%%%%%%%%%%

\end{document}